\documentclass[12pt,reqno,a4paper]{amsart}
\usepackage{amsmath,amssymb,amsfonts}
\usepackage[mathscr]{eucal}
\usepackage{graphicx}
\usepackage{amsthm, amscd}
\usepackage{hyperref}
\usepackage{color}

\newtheorem{proposition}{Proposition}

\theoremstyle{definition}
\newtheorem{remark}{Remark}
\newtheorem{example}{Example}[section]
\newtheorem{definition}{Definition}

\newcommand{\LL}{\mathbf{L}}
\newcommand{\R}{\mathbb{R}}

\newcommand{\K}{\mathbb{K}}

\def\a{\alpha}
\def\b{\beta}

\def\be{\begin{equation}}
\def\ee{\end{equation}}
\def\bee{\begin{equation*}}
\def\eee{\end{equation*}}

\def\p {\partial}
\def\w {\omega}
\def\E {{\bf E}}
\def\X {{\bf X}}

\def\bs {{\bf s}}
\def\K {{\bf K}}

\def\L {{\mathcal{L}}}
\def\G{{\mathcal{G}}}
\def\F{{\mathcal{F}}}

\title[Odd symmetric tensors...]
{Odd Symmetric Tensors, and an Analogue of the Levi-Civita Connection for an Odd Symplectic Supermanifold}

\author{H.~M.~Khudaverdian}

\author{M.~Peddie}

\address{School of Mathematics,  University of Manchester,
 Oxford Road,  Manchester   M13 9PL,  UK}
\email{khudian@manchester.ac.uk\\matthew.peddie@manchester.ac.uk }



\begin{document}

\maketitle

\begin{abstract}
We consider odd Poisson (odd symplectic) structure
on supermanifolds induced by an odd symmetric rank $2$ (non-degenerate)
contravariant tensor field. We describe the
  difference between odd Riemannian and odd symplectic structure
in terms of the Cartan prolongation of the corresponding Lie algebras,
and formulate an analogue of the Levi-Civita theorem
 for an odd symplectic supermanifold.
\end{abstract}

\section{Introduction}
The role of an odd symmetric rank $2$ contravariant tensor field
on a supermanifold is twofold. On one hand it
may be considered as the principal symbol of an odd second order
differential operator; on the other, the tensor field
defines an odd
bracket on functions, an odd Poisson bracket if the bracket
 satisfies a Jacobi identity. It is very illuminating to utilise
 this versatility to study the properties of odd brackets in terms
 of odd second order operators (see \cite{KhVor1, KhVor2}).

The fact that the study of odd brackets and odd operators may be
combined does not have an analogy in usual mathematics (where by
`usual' we mean without anticommuting variables). In the usual
setting, symmetric tensors are related with Riemannian geometry and
second order operators, whereas antisymmetric tensors are related
with Poisson and symplectic geometry. In supermathematics the notion
of symmetry for tensors becomes more subtle and the difference between
odd Riemannian and odd symplectic structures has to be based on the
fact that a Riemannian structure is ``rigid'', whilst a symplectic
structure is ``soft'', i.e. the space of infinitesimal isometries
(the Killing vector fields) of a Riemannian structure is finite dimensional,
and the space of infinitesimal isometries (the Hamiltonian vector fields)
of a symplectic structure is infinite dimensional. This fundamental
difference is a consequence of the difference in the Cartan prolongation
of the corresponding Lie algebras. When anticommuting variables are present,
it is this characteristic feature that must be used to distinguish between these two
structures. We discuss this phenomenon in the next section.

In the last section we study properties of a second order compensating field
which naturally arises on odd Poisson supermanifolds.
In the case when the Poisson structure is non-degenerate
(an odd symplectic structure), this compensating field may be
defined uniquely in a way analogous to the Levi-Civita connection.

\section{What Distinguishes Riemannian and Symplectic Structure.}
Recall the following textbook facts: Riemannian geometry and
second order operators are yielded by symmetric rank $2$ contravariant
tensor fields, and Poisson and symplectic geometry arise from antisymmetric
tensors. In more detail; a symmetric tensor field
$G = g^{ik}(x)\p_k\otimes \p_i$
($g^{ik}(x)=g^{ki}(x)$) can be considered as the principal symbol of a
second order operator $\Delta = \frac{1}{2}g^{ik}(x)\p_k\p_i + \cdots$,
and, if the tensor is non-degenerate and positive-definite,
defines a Riemannian
metric $g_{ik}(x)dx^k\otimes dx^i$. A rank $2$ antisymmetric tensor field
$P = P^{ik}(x)\p_k\wedge\p_i$ ($P^{ik}=-P^{ki}$) defines a bracket on
functions on the manifold $M$: $\{f,g\} = -\{g,f\}=\p_ifP^{ik}\p_kg$,
which is a
Poisson bracket if $P$ obeys the Jacobi identity:
$P^{ir}\p_r P^{jk}+\hbox{cyclic permutations}=0$.
If the Poisson structure is non-degenerate,
the inverse tensor field $\w_{ik}(x)dx^k\wedge dx^i$
($\w_{ik}={(P^{-1})}_{ik}$)
defines a symplectic structure on $M$.

What happens in the case when $M$ is a supermanifold? In the same way one
can consider a symmetric rank $2$ contravariant tensor field,
              \bee
     \E=E^{AB}(z){\p\over \p z^B}\otimes{\p\over \p z^A}\,,\quad
               E^{BA}=(-1)^{p(B)p(A)}E^{AB}\,,
                   \eee
which may be considered as the principal symbol of a second order operator
                   \bee
  \Delta={1\over 2}E^{AB}(z){\p\over \p z^B}{\p\over \p z^A}+\cdots\,.
                     \eee
We choose local coordinates $z^A=(x^a,\theta^\a)$ on $M$, where
$x^a$ are even (bosonic) coordinates, their parity $p$ is $p(x^a)=0$,
 and $\theta^\a$ are odd (fermionic) coordinates
with parity $p(\theta^\a)=1$. Odd coordinates
anticommute:
             \bee
x^ax^b=x^bx^a,\quad x^a\theta^\a=\theta^\a x^a,\quad {\rm whereas}\quad
 \theta^\a\theta^\b=-\theta^\b\theta^\a\,.
             \eee
Respectively for derivatives,
                \be\label{signrule1}
 {\p\over \p z^A}{\p\over \p z^B}=(-1)^{p(A)p(B)}
            {\p\over \p z^B}{\p\over \p z^A}\,,
               \ee
where we denote by $p(A)$ the parity of coordinate $z^A=(x^a,\theta^\a)$.

 We  have to distinguish two cases: when the tensor field
$\E=E^{AB}\p_B\otimes\p_A$
 is even and when it is odd. Consider for example
a $p|q\times p|q$ matrix
                 $
       \begin {pmatrix}
          A& 0\cr 0& B
       \end {pmatrix}
                 $,
where $A$ is a $p\times p$ matrix and $B$ is a $q\times q$ matrix, both containing
even entries. A tensor field defined by this matrix is an even tensor field on
$p|q$-dimensional superspace $\R^{p|q}$. It is a symmetric field  if $A$ is
symmetric and $B$ is antisymmetric, and vice versa, the tensor field is
antisymmetric if $A$ is an antisymmetric matrix and $B$ is a symmetric matrix.

Now let $K$ and $L$ be two $n\times n$ matrices with even entries. Then a
tensor field defined by the $n|n\times n|n$ matrix
           $
       \begin {pmatrix}
          0& K\cr L& 0
       \end {pmatrix}
                 $
is an example of an odd tensor field. This field is symmetric (antisymmetric)
if $K=L$ ($K=-L$). An important case of this is the following: for the $n\times  n$
 unity matrix $I$, consider two $n|n\times n|n$ matrices:
                   \be\label{butin}
                             S=
       \begin {pmatrix}
           0 & I\cr I& 0
       \end {pmatrix},\qquad
     \   G=\begin {pmatrix}
           0 & I\cr -I& 0
       \end {pmatrix}.
                  \ee
Matrix $S$ defines an odd symmetric contravariant tensor field, whilst
 matrix $G$ defines an odd antisymmetric contravariant tensor field. Later
  we will see that $S$ leads to odd symplectic geometry and $G$ to odd Riemannian geometry.

In the same way as for usual manifolds, an even symmetric tensor field
 yields Riemannian structure
and an even antisymmetric tensor field obeying the Jacobi identity yields
Poisson structure (see for details \cite{KhVor1}).
This is not the case for odd tensors where the notion of symmetry becomes
more subtle.

\smallskip

 {\bf Statement:}
{\it An odd contravariant symmetric tensor field (obeying the Jacobi identity)
defines an odd Poisson structure.
An odd contravariant antisymmetric tensor field obeying a non-degeneracy
condition defines an odd Riemannian structure.}

Discuss this statement.

\begin{example}\label{odd}
Consider the $n|n$-dimensional superspace $\R^{n|n}$ with coordinates
$(x^1,\dots,x^n|\theta_1,\dots,\theta_n)$, together with the odd symmetric
rank $2$ contravariant tensor field defined by the matrix $S$ in equation \eqref{butin}.
The components of $S$ are constants and hence the Jacobi identity is
fulfilled. We come to an odd non-degenerate Poisson bracket (symplectic structure)
 defined by the relations
         \bee
  \{x^i,\theta_j\}=\delta^i_{j}\quad {\rm and}\quad
  \{x^i,x^k\}=\{\theta_i,\theta_k\}=0\,.
       \eee
(The coordinates $x^i,\theta_j$ are Darboux coordinates
for this symplectic structure.)  
For arbitrary functions,
    \bee
    \{f,g\}=
   {\p f\over \p x^i}{\p g\over \p \theta_i}+
        (-1)^{p(f)}{\p f\over \p \theta_i}
        {\p g\over \p x^i}\,.
    \eee
This is the well-known Batalin-Vilkovisky odd bracket \cite{BV}.

\end{example}

 Notice that an odd Poisson 
bracket is symmetric, that is
    $
    \{f,g\} = (-1)^{p(f)p(g) + p(f) + p(g)}\{g,f\}.
    $
What is the reason for this paradoxical change in symmetry?

  In usual mathematics,
 given a non-degenerate (anti)symmetric tensor,
its inverse tensor is also
  (anti)symmetric. This is no longer the case when passing to
the super setting
where the symmetry of the inverse tensor is now dependent on the parity.

\begin{example}\label{mistery}
Consider a rank $2$ contravariant tensor field $L^{AB}$ with $L_{AB}$ its
inverse: $L^{AB}L_{BC} = \delta^A_C$ (if it exists).
Then one sees that
    \begin{equation}\label{inverses}
    L^{AB} = \pm(-1)^{p(A)p(B)}L^{BA}\quad\Rightarrow\quad L_{AB} =
\mp(-1)^{(p(A)+1)(p(B)+1)+p(L)}L_{BA}\,.
    \end{equation}
Consider also the map
 of tensors $X\mapsto \tilde{X}$,
    \begin{equation}\label{shift in parity}
    X^{AB}\mapsto \tilde{X}^{AB} = (-1)^{p(A)}X^{AB}\,.
        \end{equation}
If $X^{AB}=\pm (-1)^{p(B)p(A)}X^{BA}$ then
  $\tilde{X}^{AB}=\mp (-1)^{(p(B)+1)(p(A)+1)}\tilde{X}^{BA}$.
For example, if $L$ is an odd non-degenerate
symmetric (antisymmetric) tensor,
then its inverse is also symmetric (antisymmetric) but
\emph{with respect to a shift in parity}, and the inverse
to the tensor $\tilde{L}$ is an odd antisymmetric (symmetric)
tensor with respect to usual parity.
   \end{example}
This symmetry shift can be explained by the parity reversal functor
  $\Pi:V\rightarrow \Pi V$
 (which reverses the parity in a vector space $V$).
 This functor defines a
canonical isomorphism $V\otimes V\to\Pi V\otimes\Pi V$
(see equation \eqref{shift in parity}), which
induces a canonical isomorphism $S^2(\Pi V)\cong\Pi^2\wedge^2 V$
between the symmetric square of $\Pi V$ and the wedge square
of $V$. When $L$ is odd, $L$ defines an isomorphism between
$T^*M$ and $\Pi TM$ which induces a shift of symmetry.
 (For details see the appendix of the article \cite{Vorvolume1}.)

 Symmetry and antisymmetry cease to be the ``wall'' between odd
symplectic and odd Riemannian geometry. In order to distinguish between the
 structures one has to consider  other differences. Recall:
 Riemannian geometry possesses only a finite dimensional space of infinitesimal
 isometries (Killing vector fields), and symplectic geometry possesses an
 infinite dimensional space of infinitesimal isometries (Hamiltonian vector
 fields; each induced from a Hamiltonian function). Algebraically, this difference
 is expressed in terms of the difference in the Cartan prolongation of the
 orthogonal and symplectic Lie algebras.
\begin{definition}
Let $\G$ be a subalgebra of the linear Lie algebra $gl(n,\R)$. The $k$th
Cartan prolongation of $\G$ is the space $\G_k$ ($k=0,1,2,\dots$) of symmetric
$k+1$-linear maps $l:\R^n\times\dots\times \R^n\to \R^n$ such that for every
$k$ vectors $v_1,\dots,v_k$, the linear map
    \bee
    \R^n\ni v\to l(v,v_1,\dots,v_k)\in \R^n
    \eee
belongs to the Lie algebra $\G$ (see e.g. \cite{K1}). In components,
 elements of $\G_k$ are tensors $T^i_{jm_1\dots m_k}$ of type $\begin{pmatrix}1\cr k+1\end{pmatrix}$
 which are symmetric over all lower indices and, for all fixed values of
  $m_1,\ldots,m_k$, $T^i_{jm_1\dots m_k}$ belongs to the Lie algebra $\G$.
\end{definition}
The following textbook example illustrates the relation between the Cartan
prolongation of the space of infinitesimal isometries and the rigidity of the structures.

\begin{example}\label{killing example}
Let $\mathbb{R}^{2n}$ have Cartesian coordinates $(x^i)$ and consider the
$2n\times 2n$ unity matrix $I=I_{2n}$
and the antisymmetric matrix $J = \begin{pmatrix} 0&I_n\\-I_n&0\end{pmatrix}$.
The unity matrix
defines the standard Euclidean metric $G = \sum^{2n}_{i=1}(dx^i)^2$, whilst
 the matrix $J$ defines
a symplectic structure (in Darboux coordinates)
$\w = \sum^{n}_{i=1}dx^i\wedge dx^{i+n}$.

Let $\K = K^i(x)\p_i$ be a Killing vector field preserving the metric $G$
(an infinitesimal isometry):
    \begin{equation}\label{rieman killing}
    \L_\K G = 0,\quad \mbox{i.e.}\quad \frac{\p K^i(x)}{\p x^j}
     + \frac{\p K^j(x)}{\p x^i} = 0,
    \end{equation}
where $\L$ is the Lie derivative. Similarly, let $\LL$ be a vector field
preserving the symplectic structure $\omega$:
    \begin{equation}\label{symp killing}
    \L_\LL\omega = 0,\quad \mbox{i.e.}\quad \frac{\p L^m(x)}{\p x^j}J_{mi}
     + J_{jm}\frac{\p L^m(x)}{\p x^i} = 0.
    \end{equation}
If we differentiate equation \eqref{rieman killing} by any coordinate $x^k$ we come to
    \begin{equation}\label{cartanorthogonal1}
    T^i_{kj}+T^j_{ki}=0,\quad\mbox{ where } \quad T^i_{kj} = \frac{\p^2K^i}{\p x^k\p x^j}.
    \end{equation}
The tensor $T^i_{kj}$ is symmetric
with respect to the lower indices $k$ and $j$,
and antisymmetric in indices $j$ and $i$ by \eqref{cartanorthogonal1}. Hence this
tensor vanishes:
    \begin{equation}\label{cartanorthogonal2}
    T^i_{kj} = -T^j_{ki} = -T^j_{ik} = T^k_{ij} = T^k_{ji} = -T^i_{jk} = -T^i_{kj}.
    \end{equation}
Since $T^i_{kj}\equiv 0$, we see that
 $K^i=c^i+B^i_jx^j$ and so all infinitesimal
isometries of the Euclidean metric are translations
and infinitesimal rotations.
 Now notice that equation \eqref{cartanorthogonal1}
reads that the tensor $T^i_{kj}$
 belongs to $\mathfrak{so}_1(n)$, the first Cartan prolongation of the special
 orthogonal algebra $\mathfrak{so}(n)$, and equation \eqref{cartanorthogonal2}
 reads that the first Cartan prolongation of
the algebra $\mathfrak{so}(n)$ vanishes.
 This algebraic fact explains the rigidity of Riemannian geometry.

In the symplectic case,
equation \eqref{symp killing} can be rewritten as
    \begin{equation}\label{symp killing2}
    {\p L_i\over \p  x^j}={\p L_j\over \p  x^i},\quad
{\rm where}\quad  L_i=L^m J_{mi},\quad{\rm since}\quad J_{im}=-J_{mi}\,.
    \end{equation}
We see that this equation, contrary to equation \eqref{cartanorthogonal1},
has an infinite dimensional space of solutions. Every function $\Phi$
(a Hamiltonian function) defines $L_i=\p_i\Phi(x)$
which is a solution of equation \eqref{symp killing2} (respectively
every Hamiltonian vector field $L^i\p_i=J^{ij}\p_j\Phi\p_i$ is a solution
of equation \eqref{symp killing}).   In other words,
all Cartan prolongations $\mathfrak{sp}_k(n)$
of the symplectic Lie algebra $\mathfrak{sp}(n)$
are non-trivial. An arbitrary
rank $k+2$ symmetric tensor $L_{m_{1}\cdots m_{k}jr}$
defines a tensor $L^i_{m_{1}\cdots m_{k}j} =
L_{m_{1}\cdots m_{k}jr}J^{ri}$
belonging to the $k$th prolongation of the symplectic algebra.
\end{example}
Because of the subtleties with the symmetry of odd tensors, this example
suggests that the distinguishing feature between odd Riemannian and odd
symplectic structure should be the rigidity of the structures described in
terms of the Cartan prolongation of the corresponding Lie algebras.

\begin{example}
Consider the superspace $\R^{n|n}$ together with the odd symmetric tensor field
 $\E_{odd\,symp.}$ defined by the symmetric matrix $S$ in equation
\eqref{butin}, and the odd antisymmetric tensor field $\E_{odd\,riem.}$
defined by the antisymmetric matrix $G$ in the same equation \eqref{butin}.

 One can show in a similar way to Example \ref{killing example}
that the space of vector fields preserving the symplectic structure
$\E_{odd\,symp.}$ is infinite dimensional, and the space of
vector fields preserving the Riemannian structure is finite dimensional.
  Indeed, the first Cartan prolongation of the Lie algebra of vector
  fields preserving $\E_{odd\,riem.}$
vanishes. Namely, if we denote by $\K'$ a vector field
which preserves the tensor
field $\E_{odd\,riem.}$ (compare with equation \eqref{rieman killing}), then
we come to the equation
                      \bee
     T'^{A}_{CB}=-T'^B_{CA}(-1)^{p(B)p(A)},\quad\mbox{ for }\quad
      T'^{A}_{CB}={\p^2 K'^A\over \p z^C z^B}.
                     \eee
Compare it with equation \eqref{cartanorthogonal1}. This equation reads that
the tensor $T'^{A}_{BC}$ belongs to the first Cartan prolongation of the Lie
algebra of vector fields preserving the odd tensor
field $\E_{odd\,riem.}$, and
it vanishes identically in the same way as
its counterpart in equation \eqref{cartanorthogonal2}.

   For the odd symplectic structure $\E_{odd\,symp.}$
the conclusions are analogous to
those for the symplectic structure in Example \ref{killing example}.
Every rank $k+2$ symmetric tensor
defines an element in the $k$th Cartan prolongation of the
Lie algebra of vector fields preserving the odd symplectic structure
 $\E_{odd\,symp.}$. Equivalently, every Hamiltonian function $\Phi(x)$
defines a Hamiltonian vector field preserving the odd symplectic structure.
\end{example}

\begin{remark}
  We would like to note article \cite{Khjmp1}, in which
  vector fields which simultaneously preserve both even and odd
non-degenerate Poisson brackets were studied. It was shown in this paper
that the space of these
vector fields is finite dimensional. This fact was deduced from
considerations which implicitly involved the calculation of
Cartan prolongations (the vanishing of the first Cartan prolongation
of the Lie algebra of vector fields preserving both even and odd brackets).
\end{remark}

\section{Odd Second Order Operators and Odd Poisson Structure.}
Return now to an odd symmetric tensor field
$\E=E^{AB}(z)\p_B\otimes\p_A$ defined on a supermanifold $M$. To begin, we will
 briefly recall some of the results from the article \cite{KhPed1}. Denote
  by $\F_\E$ the class of odd second order self-adjoint differential
  operators with the principal
  symbol $\E$ acting on half-densities on this supermanifold $M$,
    \be\label{operatordelta}
    \F_\E\in \Delta\colon\quad
    \Delta=\frac{1}{2}\left(E^{AB}(z)\p_B\p_A+
       \p_BE^{BA}(z)\p_A+U(z)\right)\,,
                       \ee
(where $U(z)$ is an odd function on $M$, $p(U)$=1).
  If $\bs=s(z)\sqrt{Dz}$ is a half-density, then
                \bee
   \Delta\bs=\frac{1}{2}\left(\p_B\left(E^{BA}\p_A s(z)\right)
    + U(z)s(z)\right)\sqrt{Dz}.
               \eee
Any two operators in $\F_\E$ differ
by an odd scalar function (see for detail \cite{KhPed1}).

   The term $U = U(z)$ is called the potential field, and
 transforms under a change of local
coordinates in the following way:
    \begin{equation}\label{transform}
    U' = U + \frac{1}{2}\p_{A'}\left(E^{A'B'}\p_{B'}\log J\right)
    + \frac{1}{4}\left(\p_{A'}\log JE^{A'B'}\p_{B'}\log J\right),
    \end{equation}
where $J$ is the Berezinian (superdeterminant) of the Jacobian
of the coordinate change.
The potential field acts as a second order
compensation field (a second order connection) on
the manifold $M$; as a first order connection compensates the
action of diffeomorphisms on the first derivatives, the potential
field compensates the action on the second derivatives.

   We now consider
 the operator $\Delta^2$. Since $\Delta$ is an odd
 self-adjoint operator of order $2$,
  $\Delta^2$  is an even anti-self-adjoint operator of order
equal to either
 $3$, $1$ or else $\Delta^2=0$. The condition that $\E$ defines a
 Poisson structure and
 the relations between this structure and the class $\F_\E$
of operators can be
 summarised in the
 following statement.

\begin{proposition}\label{khudped} (\cite{KhPed1}).
  {\it Let $\Delta$ be an arbitrary operator in the class $\F_\E$.
  Then the odd symmetric tensor field $\E$
defines an odd Poisson structure on the manifold $M$, i.e.
it obeys the
Jacobi identity, if and only if the order of the operator $\Delta^2$
is equal to $1$ or else $\Delta^2=0$. In this case
 the operator $\Delta^2$ defines a vector field $\X=\X_\Delta$ such that
    \begin{equation}\label{mod vf}
    \Delta^2=\L_\X\,,
    \end{equation}
  where $\L_\X$ is the Lie derivative along the vector field $\X$.
The vector field $\X=\X_\Delta$ is called the modular
vector field of the operator $\Delta$, and preserves
 the Poisson structure.
 If $\Delta'$ is another arbitrary operator from the class $\F_\E$,
that is, $\Delta'=\Delta+F$, then
         $
    \X_{\Delta'}=\X_{\Delta}+D_F,
         $
where $D_F$ is the even Hamiltonian vector field corresponding
to the odd function $F$.
   The corresponding
  equivalence  class of the modular vector field
   (in Lichnerowicz-Poisson cohomology)  is called the
   modular class of the odd Poisson manifold.}
\end{proposition}

\smallskip

If $\E$ defines an odd Poisson structure on
$M$ then the modular vector field
  $\X=\X_\Delta$ of operator \eqref{operatordelta}
has the following local appearance:
    \be\label{mod vf local}
    \X =\frac{1}{2}\p_C\left(E^{CD}\p_D\p_BE^{BA}\right)\p_A
     + (-1)^{p(A)}E^{AB}\p_BU\p_A.
    \ee

  Assume now that the tensor field
$\E$ defines an odd non-degenerate Poisson structure
(an odd symplectic structure) on the supermanifold $M$.
 In Darboux coordinates $(x^i,\theta_j)$ (see example \ref{odd}),
one may naively define the canonical operator $\Delta$
on half-densities by
    \begin{equation}\label{canonical lap}
    \Delta\bs = \frac{\p^2s(x,\theta)}{\p x^i\p \theta_i}\sqrt{D(x,\theta)}.
    \end{equation}
What is remarkable is that this local expression defines
  an operator globally on $M$ \cite{Khcmp2};
in arbitrary Darboux coordinates it has the same appearance.
   In other words, the vanishing of the potential $U$
in one set of Darboux coordinates implies that it vanishes in arbitrary Darboux
coordinates. We see that
 if $\E$ defines a non-degenerate odd Poisson structure, then
the class $\F_\E$ possesses a distinguished operator, the
  odd canonical operator \eqref{canonical lap} defined by the condition
            \be\label{nooddconstant}
             U=0\quad\hbox{in Darboux coordinates}\,.
          \ee
The expression for the potential of the odd canonical operator
  was calculated in arbitrary coordinates in \cite{Bering1}.
 It has the appearance
    \begin{equation}\label{unique exp}
    U(z) = \frac{1}{4}\p_B\p_A E^{AB}(z) -
    (-1)^{p(B)(p(D)+1)}\frac{1}{12}\p_AE^{BC}(z)E_{CD}(z)\p_BE^{DA}(z)\,,
    \end{equation}
where $E_{AB}$ is the inverse tensor to $E^{AB}$.

\begin{proposition}
{\it For an odd symplectic supermanifold there exists a unique
potential field $U$ defining the odd canonical operator \eqref{canonical lap}.
This potential field vanishes in arbitrary Darboux coordinates.
In arbitrary local coordinates it is given by expression \eqref{unique exp}.
The potential $U$
acts as a second order compensating field and transforms under a
change of coordinates according to \eqref{transform}.}
\end{proposition}

This proposition can be considered as a far analogue to the Riemannian case
of the unique first order
   compensating field -- the Levi-Civita connection.

Notice that the modular vector field \eqref{mod vf}
of the odd canonical operator vanishes.
(In particular, this means that the modular class of an odd symplectic manifold
vanishes.)
The vanishing of the modular vector field of the canonical operator
means that the potential $U$ in \eqref{unique exp} is a solution of the
 first order differential equations,
     $$
  \frac{1}{2}\p_C\left(E^{CD}\p_D\p_BE^{BA}\right)
     + (-1)^{p(A)}E^{AB}\p_BU = 0\,,
   $$
which follow from equation \eqref{mod vf local}.
The solution of these equations is unique up to an odd constant.
Condition \eqref{nooddconstant} implies that this odd constant
vanishes for the canonical operator \eqref{canonical lap}. (See also article \cite{BatBer1}).

\begin{remark}
We would like to note that we currently have no
conceptually clear way of deriving formula \eqref{unique exp}.
On the other hand, the
vector field \eqref{mod vf local} has some mysterious
properties which may be part of a calculus for odd
Poisson manifolds and in particular for odd symplectic
geometry. We think that understanding these properties
 will elucidate the geometrical structure of
expression \eqref{unique exp}.
 This is a work in progress.
\end{remark}

\section*{Acknowledgment}
We are grateful to Th. Voronov for encouraging discussions.

\end{document}